# Thermoelectric properties of Half-Metallic FeMnScGa Using First Principle Calculation

Shubham Singh,[1)] Saurabh Singh,[2)] Nitinkumar Bijewar,[1)] and Ashish Kumar [3,b)]

[1](*Department of Physics, university of Mumbai, kalian Campus, Santacruz (E)-400098, Mumbai, India.*)
[2](*Toyota Technological Institute, Hisakata, 2-12-1, Tempaku-ku, Nagoya 468-8511, Japan.*)
[3](*Inter-University Accelerator Centre, Aruna Asaf Ali marg, Vasant Kunj, New Delhi-110067, India.*)

[a)]Corresponding author: saurabhsingh@toyota-ti.ac.jp
[b)]Corresponding author: ashish@iuac.res.in

**Abstract.** Here, we have investigated the thermoelectric properties of FeMnScGa alloy by combined use of *full potential linearized augmented plane-wave* (FP-LAPW) method and Boltzmann transport theory implemented in Wien2K and BoltzTraP code, respectively. Using the TB-mBJ potential, half-metallic nature is clearly observed with energy band gap of ~0.41 eV for down spin channel. Transport coefficients in 200-1000 K temperature range are investigated under the constant relaxation time approximations ($\tau=10^{-14}$ s). Total contributions to the Seebeck coefficients from up and dn-spin channels are estimated by using two-current model. The calculated values of Seebeck coefficients are found to be negative in the entire temperature region under present study, which suggests the *n*-type characteristic of this alloy. The magnitude of Seebeck coefficient is found to be increasing with temperature and exhibit the large value ~ -87 μV/K in 700-1000 K temperature range. The maximum in magnitude of total electrical conductivity value is found to be ~2.0 × $10^5$ $\Omega^{-1}m^{-1}$ at 1000 K. The present study shows, with half-metallic electronic structure, FeMnScGa have very good thermoelectric behavior, and can be suitable for thermoelectric application in high temperature region.

## INTRODUCTION

In the past several decades, heusler alloy have been extensively explored both theoretically as well as experimentally due to their fascinating applications in the field of spintronic based electronic devices. Based on the crystal structure, heusler alloys are generally classified into three different categories named as half-heusler, full-heusler, and quaternary heusler alloy (QHA) [1,2]. Among these class of maerials, a detail investigation on QHA have been recently attracted much attention, as they are found to exhibit half-metallic and insulating electronic structure, which are easily tunable with their crystallographic degree of freedoms by the constituent elements in the unit cell. The typical formula of QHA is XX'YZ, where X, X' and Y are transition metal elements, and Z is main group element. The most common crystal structure is LiMgPdSn characterized by the space group *F-43m* (no. 216) [3]. The major investigations so far are mainly focused on the spin driven properties in these type of materials. In addition to spintronic applications, thermal transport properties are also found to be interesting for the energy conversion from waste heat to electrical energy through the mechanism of thermoelectricity. The efficiency of the materials to convert heat into electric energy depends upon the *figure-of-merit (ZT)*, which can be expressed as: ZT = ($\alpha^2\sigma$)T/$\kappa$, where $\alpha$, $\sigma$, $\kappa$ and T are Seebeck coefficient, electrical conductivity, electronic thermal conductivity and absolute temperature, respectively [4]. The expression of *ZT* suggest that a material having higher value of $\alpha$ and $\sigma$, along with lower $\kappa$ value is very good to attain the large *figure-of-merit*, thus suitable for good thermoelectric applications. In the search of better material with such selection criteria, material with half-metallic electronic structure are found to be one of the possible candidate as up-spin channel have the metallic characteristics give high electrical conductivity, whereas dn-spin channel with narrow band gap possesses large Seebeck coefficient, thus together a large power factor ($\alpha^2\sigma$) are achievable.

Recently, a new class of QHA material FeMnScGa is theoretically investigated by Gao et al, which shows the half-metallic characteristics, and based upon the formation energy calculations proposed for the spintronic application [5]. In order to find the applicability of this material for thermoelectric applications, a detailed investigations of transport properties become necessary. Thus, in the present work, we have done the geometrical optimization, electronic structure calculations using density functional theory along with the transport coefficient calculations in 200-1000 K temperature range using Boltzmann transport theory based computational tool. Two-current model is used to calculate the total Seebeck coefficient, and obtained results have been discussed for the consideration of FeMnScGa alloy as one possible material for high temperature thermoelectric applications.

## COPUTATIONAL DETAILS

The geometrical optimization and electronic structure calculations were carried out by using the *full potential linearized augmented plane-wave* (FP-LAPW), implemented in WIEN2K code, whereas transport coefficients were estimated by using the BoltzTraP code [6,7]. For the exchange correlation functional, we have chosen PBE and TB-mBJ potential [8,9]. For the self-consistency field calculation, the energy convergence criteria were set to $10^{-6}$ Ry. The k-mesh values for geometrical optimization, electronic structure, and transport coefficients calculations were set to 10×10×10, 30×30×30, and 50×50×50, respectively. The Wyckoff positions of constituent elements in FeMnScGa are Fe(1/4,1/4,1/4), Mn(3/4,3/4,3/4), Sc(1/2,1/2,1/2), and Ga(0,0,0).

## REULTS AND DISCUSSIONS

The unit cell of FeMnScGa is shown in Fig. 1a, whereas energy vs volume plot obtained from the geometrical optimization is shown Fig. 1b. The optimized lattice parameter is estimated by equation of states fit using Birch Murnaghan expression. With PBE exchange correlation functional, the value of optimized lattice parameter and bulk modulus are found to be 6.10 angstrom and 120.41 GPa, respectively.

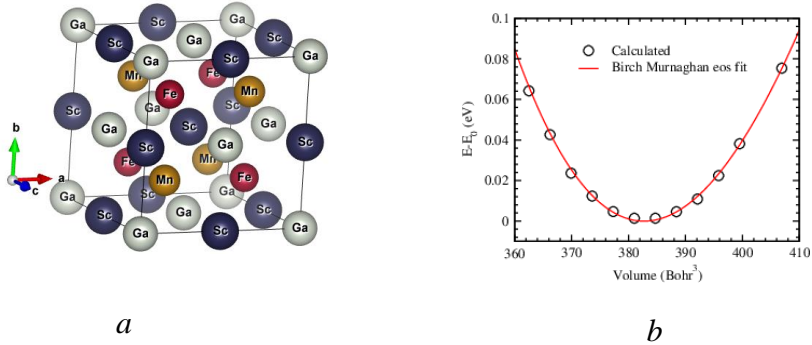

*a*            *b*

Fig 1. (*a*) Unit cell of FeMnScGa, and (*b*) Energy vs Volume plot.

Total density of states plot estimated using PBE and TB-mBJ potential for up and dn-spin are plotted together in Fig. 2. It is clearly evident from TDOS plot, the use of PBE exchange correlation give the metallic characteristic. However, the number of total valance charge in the unit cell is 21 electrons, and according to the Slater-Pauling rule (M = Nv – 24), we can find the M value equal to 3 [10]. Thus, with an integer value of M, the half-metallic characteristics from the TDOS plot is expected for this alloy. Generally, PBE under estimate the band gap, and unable to produce a gap in case of narrow band gap. While using the TB-mBJ, which is found to calculate the electronic structure more accurately in comparison to PBE, we clearly observe the half-metallic characteristic for FeMnScGa alloy. The up-spin channel has finite density of states at Fermi level (E = 0 eV) showing metallic characteristics, whereas appearance of energy band gap of ~0.410 eV in dn-spin channel suggests the insulating nature. The calculated value of total spin magnetic moment per unit cell is found to be 3.0 μB, which is an integer number, and confirms the half metallic characteristic of this alloy. Also, the sharp edge in conduction band for dn-spin channel close to Fermi level is suitable for large Seebeck coefficient, and also suggests the *n*-type characteristic of this material. Both valance band (VB) and conduction band (CB), have large density of states about fermi level within 1 eV range, which is very important feature to obtained the better transport property. At finite temperature, fraction of thermally excited electrons from top of the VB of dn-spin channel will be accommodated in the available states in CB, and have the contribution in transport coefficients. For the up-spin channel, the large number of electron will contribute in the electrical conductivity.

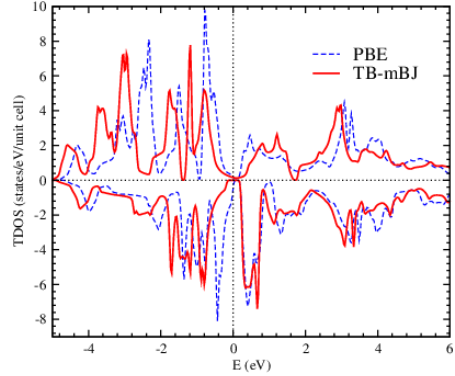

Fig. 2 Total density of states plot for FeMnScGa, PBE (blue dotted line) and TB-mBJ (red solid line).

Dispersion curve calculated (for up and dn-spin channel) along the high symmetric k-point is shown in Fig. 3. It is clearly observed that bands are crossing across the fermi level for the up-spin channel, whereas for dn-spin channel, indirect band gap of ~0.41 eV is found. The bands crossing at Fermi level have the large number of free electron and mainly contribute to the electrical conductivity. For dn-spin channel, top of VB at $\Gamma$ point is triply degenerate, whereas bottom of the CB at **L** point is doubly degenerate, and mainly responsible for the contributions in Seebeck and electrical conductivity when the thermal heat gradient will be applied.

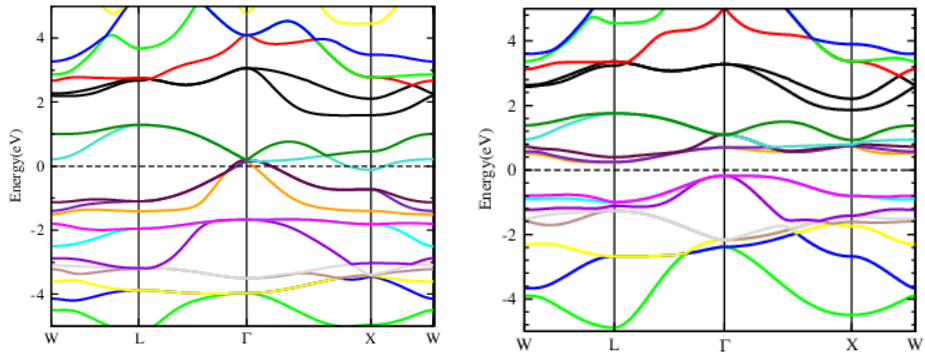

Fig. 3 Band structure plot for FeMnScGa along **W-L-$\Gamma$-X-W**, up-spin (left) and dn-spin (right).

Thermal transport coefficients as a function of temperature are estimated in the 200 K-1000 K range, which are plotted in Fig. 4. With metallic characteristic, up channel shows small Seebeck coefficients, whereas have the large electrical conductivity. Here, the value of $\sigma$ is calculated by taking the constant value of relaxation time i.e. $\tau = 10^{-14}$ s. For the dn-spin channel, a large Seebeck coefficient is obtained at 200 K, due to the small band gap in minority spin. The temperature dependent of $\sigma$ for dn-spin channel shows a typical semiconducting behavior and increases with increase in temperature. The value of electronic thermal conductivity for both up-and dn-spin channel show increasing behavior with temperature. To get the total Seebeck coefficient, we have used two-current model expression i.e. $S_{tot} = (S\uparrow\sigma\uparrow + S\downarrow\sigma\downarrow)/(\sigma\uparrow + \sigma\downarrow)$, where up ($\uparrow$) and down ($\downarrow$) arrows represent the values of Seebeck coefficient and electrical conductivity corresponding to the up and dn-spin charge carriers, respectively [11]. In the entire temperature range under investigation we observed negative Seebeck coefficient, suggest the *n*-type material. At 200 K, the value of $S_{tot}$ is equal to -23 µV/K. With temperature, magnitude of $S_{tot}$ increases up to 700 K, and further it remains almost independent of temperature. The saturated value of $S_{tot}$ is found to be ~ -87 µV/K in 700-1000 K temperature range, which is relatively good magnitude at such high temperature. Here, an important point need to be noticed that, $S_{tot}$ is calculated by taking the constant value of relaxation time for both up and dn-spin channel charge carriers. The values of τ for up and dn-spin can be different with temperature. Also, the total electrical conductivity value can be estimated more realistic with accurate value of τ, which can be obtained with experimental measured value of electrical conductivity [11]. Thus, for further investigation of more accurate values are τ and verify the reported Seebeck coefficient, an experimental study on this compound is highly desirable. The calculation of formation energy is found to be negative, thus suggesting that this alloy is experimentally can be synthesized in laboratory [5]. The large magnitude of S and σ reported based on the first principle calculations suggests that this alloy can be a suitable materials for thermoelectric applications in high temperature range.

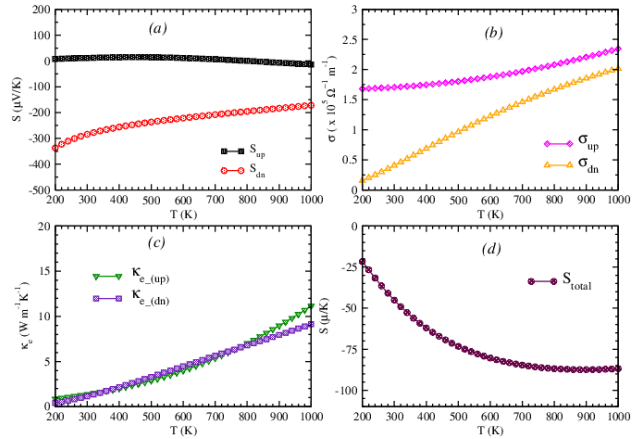

Fig. 4 Shown are (a) Seebeck coefficient (*S*), (b) electrical conductivity (σ), (c) electronic thermal conductivity ($\kappa_e$) and (d) $S_{total}$ as a function of temperature for FeMnScGa.

## CONCLUSION

In conclusion, we have systematically investigated the electronic structure and thermoelectric property of FeMnScGa alloy using first principle method. Calculation by using TB-mBJ potential give the half-metallic electronic structure of this alloy, with energy band gap of ~0.41 eV for minority charge carriers. The total magnetic moment per unit cell is found to be 3.0 μB. Using two-current model, estimated values of Seebeck coefficients in temperature range of 200-1000 K are negative, which suggests that *n*-type charge carriers have major contributions in the transport properties. The large Seebeck coefficient is found to be ~ -87 μV/K in 700-1000 K range. With large magnitude of Seebeck coefficient together with electrical conductivity, this alloy can be considering for the new thermoelectric materials in high temperature range applications.